\documentclass[twocolumn,bibnotes,showpacs]{revtex4-1}

\usepackage{graphicx}% Include Fig.~\ref{fig:overview} files
\usepackage{color}
\usepackage{soul} %needed for strikeout
\usepackage{amsmath}
\usepackage{hyperref}
\usepackage{braket}
\hypersetup{colorlinks=true, citecolor=cyan, urlcolor=blue, linkcolor=blue}

\newcommand{\RMP}[4]{\textit{#1}, Rev. Mod. Phys. \textbf{#2}, #3 (#4)}
\newcommand{\PR}[4]{\textit{#1}, Phys. Rev. \textbf{#2}, #3 (#4)}
\newcommand{\PRL}[4]{\textit{#1}, Phys. Rev. Lett. \textbf{#2}, #3 (#4)}
\newcommand{\PRR}[4]{\textit{#1}, Phys. Rev. Research. \textbf{#2}, #3 (#4)}
\newcommand{\PRA}[4]{\textit{#1}, Phys. Rev. A \textbf{#2}, #3 (#4)}
\newcommand{\PRB}[4]{\textit{#1}, Phys. Rev. B \textbf{#2}, #3 (#4)}
\newcommand{\PRApplied}[4]{\textit{#1}, Phys. Rev. Appl. \textbf{#2}, #3 (#4)}

\newcommand{\Science}[4]{\textit{#1}, Science \textbf{#2}, #3 (#4)}
\newcommand{\Nature}[4]{\textit{#1}, Nature \textbf{#2}, #3 (#4)}
\newcommand{\NPhys}[4]{\textit{#1}, Nat. Phys \textbf{#2}, #3 (#4)}
\newcommand{\NComm}[4]{\textit{#1}, Nat. Comm \textbf{#2}, #3 (#4)}
\newcommand{\SciAdv}[4]{\textit{#1}, Sci. Adv \textbf{#2}, #3 (#4)}

\begin{document}
\title{Sideband transitions in a two-mode Josephson circuit driven beyond the rotating wave approximation}
\author{Byoung-moo Ann}
\email{byoungmoo.ann@gmail.com}
\author{Wouter Kessels}
\author{Gary. A. Steele}
\affiliation{Kavli Institute of Nanoscience, Delft University of Technology, 
2628 CJ Delft, The Netherlands} 
\date{\today}

\begin{abstract}
Driving quantum systems periodically in time plays an essential role in the coherent control of quantum states.
The rotating wave approximation (RWA) is a good approximation technique for weak and nearly-resonance driven fields. However, these experiments  sometimes require large detuning and strong driving fields, for which the RWA may not hold. 
In this work, we experimentally, numerically, and analytically explore strongly driven two-mode Josephson circuits in the regime of strong driving and large detuning. 
Specifically, we investigate beam-splitter and two-mode squeezing interaction between the two modes induced by driving a two-photon sideband transition.
Using numerical simulations, we observe that the RWA is unable to correctly capture the amplitude of the sideband transition rates. We verify this finding using an analytical model that is based on perturbative corrections.
We find that the breakdown of the RWA in the regime studied does not lead to qualitatively different dynamics, but gives the same results as the RWA theory at higher drive strengths, enhancing the coupling rates compared to what one would predict. This is an interesting consequence compared to the carrier transition case, where the breakdown of the RWA results in qualitatively different time evolution of the quantum state.
Our work provides an insight into the behavior of time-periodically driven systems beyond the RWA. We also provide a robust theoretical framework for including these findings in the calculation and calibration of quantum protocols in circuit quantum electrodynamics.

\end{abstract}

\maketitle

\section{Introduction}
Time-periodic driving is a prominent technique for the $in-situ$ coherent control of quantum dynamical processes.
However, exactly solving the quantum dynamics of the system with time-varying Hamiltonian is particularly difficult \cite{Bloch-PR-1940,Giscard-PRR-2020}. As long as the drive is weak enough and nearly resonant with the quantum state transition of a target observable, then the rotating wave approximation (RWA) may provide a good estimate of the dynamics \cite{RWA,note}.
It is, however, necessary to understand the physics of quantum systems beyond the RWA from both from a fundamental and a practical perspective.
In the realm of faithful quantum information processing (QIP), the need for fast gates to suppress quantum errors requires drive strengths that could exceed those that are valid for the RWA. Motivated by this problem, theoretical effort has been directed to understand quantum driven systems beyond the RWA \cite{Bloch-PR-1940,Giscard-PRR-2020}, along with many experimental works that have used strong driving across many different physical systems. 
\cite{Andrews-JPB-1975,Fuchs-Science-2009,Tuorila-PRL-2015,Deng-PRL-2015,Laucht-PRB-2016,Pietik-PRB-2016,Sie-Science-2017,Koski-PRL-2018}. 
Although the previous studies have explored driven systems beyond the RWA, they focus on a single driven mode and do not address the coupling of different degrees of freedom (e.g., the use of driving fields to induce sideband transitions between modes).

Sideband transitions ubiquitously appear in a variety of physical systems \cite{Leek-PRB-2009,Leek-PRL-2010,Leghtas-Science, Mundhada-PRAppl-2019,Teufel-Nature-2011,Schliesser-NPhys-2008,Monroe-PRL-1995,Hennrichh-PRL-2000,Fedorov-PRA}. 
Driving systems with appropriately chosen frequencies can yield engineered interactions among different degrees of freedoms. 
To engineer a specific interaction, it is important to accurately estimate the transition rates.
In many cases the driving parameters for sideband transitions typically satisfy the requirements of the RWA, such as trapped ions, cavity-optomechanics, and Raman transitions \cite{Teufel-Nature-2011,Schliesser-NPhys-2008,Monroe-PRL-1995,Hennrichh-PRL-2000}. However, this is not always the case for the circuit quantum elecrodynamics (cQED) platform, one of the most promising QIP platforms in recent years, where a strong and far off resonant driving beyond the RWA is sometimes required \cite{BANN-PRA}. Nonetheless, current approaches to quantitative analysis still rely on the application of the RWA. When the sideband driving frequencies are far off-resonant from the transition frequencies of the system, in which the conditions for the application of the RWA should not hold, we may not currently be possible to make reliable predictions of the transition rates. 

In this paper, we study the sideband transition rates in a two-mode Josephson circuit that is induced by strong external time-periodical driving. The circuit comprises a transmon \cite{Koch-PRA-2007} that is dispersively coupled to a resonator mode. 
Specifically, we study beam splitter (BS) and two-mode squeezing (TMS) interactions between each mode, which are the simplest forms of sideband transitions in these two-mode systems. For our device, the required driving parameters are close to (TMS coupling), or far beyond (BS coupling) the RWA regime. 
We confirm a simple relationship between the transition rates and frequency shifts, which explains the data in both regimes.

We  perform  numerical simulations to support our findings. We also derive an analytical perturbation expansion that goes beyond the RWA, which is validated by our numerical results. Our findings indicate that although the RWA is clearly violated, and significantly underestimates the mode frequency shifts and the sideband transition rates for a known driving strength, the breakdown of the RWA does not result in qualitatively different behaviour but instead its effects in our measurements can be reproduced by the RWA theory using a larger drive field. Although the confirmation of a breakdown of the RWA is only possible to observe experimentally in an accurate independent calibration of the drive field, our results show the importance of including counter-rotating terms for accurate calculations of the sideband transition rates.  

\section{Theoretical description}
\label{2}
We derive an analytical expression taking a similar approach in \cite{Reagor-thesis, Leghtas-Science} but breaking the RWA. The total Hamiltonian on the lab frame is given by,
\begin{equation}
\begin{split}
     \hat{H}_{sys}^{(0)} \approx ~ \,& ({\omega}_{t}^{(0)}+\chi_{t}^{(0)}) \hat{\alpha}^\dagger \hat{\alpha} + {\omega}_{r}^{(0)}\hat{\beta}^\dagger \hat{\beta} +g(\hat{\alpha} + \hat{\alpha}^\dagger)(\hat{\beta} + \hat{\beta}^\dagger)\\
    & -\chi_{t}^{(0)}(\hat{\alpha} + \hat{\alpha}^\dagger)^{4}/12.
\label{eq0}
\end{split}
\end{equation}
Here, ${\omega}_{t}^{(0)}$ and ${\omega}_{r}^{(0)}$ are the resonant frequencies of each mode. $\hat{\alpha}$ and $\hat{\beta}$ are the mode destruction operators of the transmon and the resonator modes, respectively. $\chi_t$ is a Duffing nonlinearity of the transmon mode. $g$ is a transverse coupling between the transmon mode and the resonator mode. 
In addition to $\hat{H}_{sys}^{(0)}$, there is the driving Hamiltonian $\hat{H}_d^{(0)}=\Omega_{d}\cos{\omega_{d}t} (\hat{\alpha}+\hat{\alpha}^\dagger)$, where $\Omega_{d}$ and $\omega_{d}$ are the driving amplitude and frequency, respectively.
The total Hamiltonian $\hat{H}_{tot}^{(0)}$ is then given by $\hat{H}_{sys}^{(0)}+\hat{H}_d^{(0)}$.

It is often useful to rewrite this Hamiltonian in the normal mode basis (the normal mode annihilation operators are $\hat{a}$ and $\hat{b}$):
\begin{equation}
\begin{split}
    \hat{H}_{sys}^{(1)} \approx ~ \,& ({\omega}_{t}^{(1)}+\chi_{t}) \hat{a}^\dagger \hat{a}+ {\omega}_{r}^{(1)} \hat{b}^\dagger \hat{b}  \\
    & - \frac{1}{12} \left[ \chi_t^{1/4} (\hat{a} + \hat{a}^\dagger) + \chi_r^{1/4} (\hat{b} + \hat{b}^\dagger) \right] ^4.
\label{eq1}
\end{split}
\end{equation}
With typical circuit QED parameters, $\chi_{t}$ is approximately the same as $\chi_{t}^{(0)}$.
$\chi_r$ is the inherited Duffing nonlinearity to the resonator mode by the coupling $g$. 
In the dispersive coupling regime ($|\omega_t-\omega_{r}| \gg g$), 
$\hat{\alpha}$ in $\hat{H}_{d}$ can be approximated by $\hat{a}$ \cite{Leghtas-Science, Gely-PRA-2018}. Then, the driving Hamiltonian can be approximated to be $\hat{H}_d^{(1)} =\Omega_{d}\cos{\omega_{d}t} (\hat{a}+\hat{a}^\dagger)$.

The total Hamiltonian in the normal mode basis is then given by $\hat{H}_{tot}^{(1)}=\hat{H}_{sys}^{(1)}+\hat{H}_{d}^{(1)}$. This can be perturbatively diagonalized by taking Schrieffer–Wolff (S–W) transformation \cite{SW} $\hat{U}(t)=e^{\hat{S}}$ with an appropriate generator $\hat{S}=\xi(t)\hat{a}^\dagger(t)-\xi(t)^{*}\hat{a}$. When $\Delta \gg \chi_{t}$, we can choose $\xi(t)=\frac{\Omega_{d}}{2\Delta}e^{-i\omega_{d}t}+\frac{\Omega_{d}}{2\Sigma}e^{i\omega_{d}t}$. Here, $\Delta=\omega_{t}^{(1)}+\chi_{t}-\omega_{d}$ and $\Sigma=\omega_{t}^{(1)}+\chi_{t}+\omega_{d}$. In this work, we treat beam splitter ($\hat{a}\hat{b^\dagger}+\hat{a^\dagger}\hat{b}$) and two-mode squeezing ($\hat{a}\hat{b}+\hat{a^\dagger}\hat{b^\dagger}$) interactions induced by two-photon driving. These appear with frequency matching conditions $2\omega_{d}\approx|\omega_{t}^{(1)}\pm\omega_{r}^{(1)}|$. 

After taking Schrieffer–Wolff transformation, collecting only the original and relevant derived terms yields,
\begin{equation}
 \begin{aligned}
    \hat{H}_{tot}^{(1)} \approx ~ \,& ({\omega}_{t}^{(1)}+\delta\omega_{t}^{(1)}+\chi_{t}) \hat{a}^\dagger \hat{a}+ ({\omega}_{r}^{(1)}+\delta\omega_{r}^{(1)}) \hat{b}^\dagger \hat{b}  \\
    & - \frac{1}{12} \left[ \chi_t^{1/4} (\hat{a} + \hat{a}^\dagger) + \chi_r^{1/4} (\hat{b} + \hat{b}^\dagger) \right] ^4 + \hat{H}_{sb},
\label{eq2-1}
\end{aligned}
\end{equation}
where $\hat{H}_{sb} = \Omega_{sb}^{(1)}/2(\hat{a}\hat{b}^\dagger e^{i2\omega_dt}+\hat{a}^\dagger\hat{b}e^{-i2\omega_dt})$ when $2\omega_{d}\approx|\omega_{t}^{(1)}-\omega_{r}^{(1)}|$, and $\hat{H}_{sb} = \Omega_{sb}^{(1)}/2(\hat{a}\hat{b}e^{i2\omega_dt}+\hat{a}^\dagger\hat{b^{\dagger}e^{-i2\omega_dt}})$ when $2\omega_{d}\approx|\omega_{t}^{(1)}+\omega_{r}^{(1)}|$.
Here, $\Omega_{sb}^{(1)}$ is the interaction rate for both the BS and TMS interactions. $\delta\omega_{t}^{(1)}$, $\delta\omega_{r}^{(1)}$ and $\Omega_{sb}^{(1)}$ can be expressed by,
\begin{equation}
\begin{split}
   \delta\omega_{t}^{(1)} = ~ -\frac{1}{2}\Omega_{d}^{2}\chi_{t}\times (\frac{1}{\Delta^2} +\frac{2}{\Delta\Sigma}+\frac{1}{\Sigma^2}),\\
    \delta\omega_{r}^{(1)} = ~ -\frac{1}{2}\Omega_{d}^{2}\chi_{tr}\times (\frac{1}{\Delta^2} +\frac{2}{\Delta\Sigma}+\frac{1}{\Sigma^2}),\\
    \Omega_{sb}^{(1)} = ~ -\frac{1}{2}\Omega_{d}^{2}\chi_{t}^{3/4}\chi_{r}^{1/4}\times (\frac{1}{\Delta^2} +\frac{2}{\Delta\Sigma}+\frac{1}{\Sigma^2}).
\label{eq2}
\end{split}
\end{equation}

In the low excitation limit, the total Hamiltonian can be reduced to, 
\begin{equation}
 \begin{aligned}
    \hat{H}_{tot}^{\bf{low}} \approx ~ ({\omega}_{t}+\delta\omega_{t}) \hat{a}^\dagger \hat{a}+ ({\omega}_{r}+\delta\omega_{r}) \hat{b}^\dagger \hat{b}
    -\frac{A_{t}}{2} {\hat{a}^\dagger\hat{a}^\dagger\hat{a}\hat{a}}
    \\ - \frac{A_{r}}{2} {\hat{b}^\dagger\hat{b}^\dagger\hat{b}\hat{b}}
     - 2A_{tr}\hat{a}^\dagger\hat{a}\hat{b}^\dagger\hat{b}
    + \hat{H}_{sb}.
\label{eq3}
\end{aligned}
\end{equation}
$\omega_{t,r}$ and $A_{t,r,tr}$ ($A_{tr}\approx\sqrt{A_{t}A_{r}}$) correspond to the transition frequencies and anharmonicities that we observe in the experiments. We can obtain $A_{t,r,tr}$ by numerically diagonalizing Eq.~\ref{eq1} \cite{Lescanne-PRAppl-2019}. The difference between $A_{t,r,tr}$ and $\chi_{t,t,tr}$ is due to the off diagonal elements in Eq.~\ref{eq2-1}.

The discrepancy between $A_{t,r,tr}$ and $\chi_{t,t,tr}$ suggests that the off diagonal elements also affect the derived quantities after taking the (S–W) transform. We hereby invoke an assumption that the effects of the off diagonal terms can be captured by replacing $\chi_{t,t,tr}$ with $A_{t,r,tr}$.
This assumption leads to a conclusion that $\delta\omega_{t,r}^{(1)}$ and $\Omega_{sb}^{(1)}$ in Eq.~\ref{eq2} should be renormalized to $\delta\omega_{t,r}$ and $\Omega_{sb}$,
\begin{equation}
\begin{split}
   \delta\omega_{t,r} = -\frac{1}{2}\Omega_{d}^{2} A_{t,tr}\times (\frac{1}{\Delta^2} +\frac{2}{\Delta\Sigma}+\frac{1}{\Sigma^2}),\\
    \Omega_{sb} = -\frac{1}{2}\Omega_{d}^{2}A_{t}^{3/4}A_{r}^{1/4}\times (\frac{1}{\Delta^2} +\frac{2}{\Delta\Sigma}+\frac{1}{\Sigma^2}).
\label{eq3-1}
\end{split}
\end{equation}
We provide the supporting information for this finding in Appendix~\ref{append-5}.
When applying the RWA, they are given by,
\begin{equation}
\begin{split}
   \delta\omega_{t,r}^{\textup{(RWA)}} = ~ -\frac{1}{2}\Omega_{d}^{2} A_{t,tr}\times \frac{1}{\Delta^2},\\
    \Omega_{sb}^{\textup{(RWA)}} = ~ -\frac{1}{2}\Omega_{d}^{2}A_{t}^{3/4}A_{r}^{1/4}\times \frac{1}{\Delta^2}.
\label{eq3-2}
\end{split}
\end{equation}
It is also interesting to investigate the case where only the counter-rotating terms in $\hat{H}_d$ affect the system. In this case, the frequency shifts and sideband transition rates are given by,
\begin{equation}
\begin{split}
   \delta\omega_{t,r}^{\textup{(CR)}} = ~ -\frac{1}{2}\Omega_{d}^{2} A_{t,tr}\times \frac{1}{\Sigma^2},\\
    \Omega_{sb}^{\textup{(CR)}} = ~ -\frac{1}{2}\Omega_{d}^{2}A_{t}^{3/4}A_{r}^{1/4}\times \frac{1}{\Sigma^2}.
\label{eq3-3}
\end{split}
\end{equation}
The detailed derivation is provided in Appendix~\ref{append-theory}.
If $\delta\omega_{t}$ is known, then replacing $\Delta$ and $\Sigma$ with $\Delta+\delta\omega_{t}$ and $\Sigma +\delta\omega_{t}$ will provide a more accurate estimate.
It is worth pointing out here that many of the previous studies do not seriously distinguish between $\chi_{t,r,tr}$ and $A_{t,r,tr}$. However, the discrepancies between $\chi_{t,r,tr}$ and $A_{t,r,tr}$ are sometimes significant, depending on the system's parameters. Renormalization of $\delta\omega_{t,r}$ and $\Omega_{sb}$ is therefore of great importance for the accurate prediction of the frequency shifts and sideband transition rates.

Eq.~\ref{eq3-1} and Eq.~\ref{eq3-2} suggests that the RWA significantly underestimates $\delta\omega_{t,r}$ and $\Omega_{sb}$ when $\Delta \sim \Sigma$ but the ratios among them are identical, regardless of whether or not we use the RWA. It is also interesting to note that there is a correlation between the co- and counter-rotating terms in Eq.~\ref{eq3-1}, which makes a significant contribution to the frequency shifts and sideband transition rates.

\begin{figure}
    \centering
    \includegraphics[width=1.05\columnwidth]{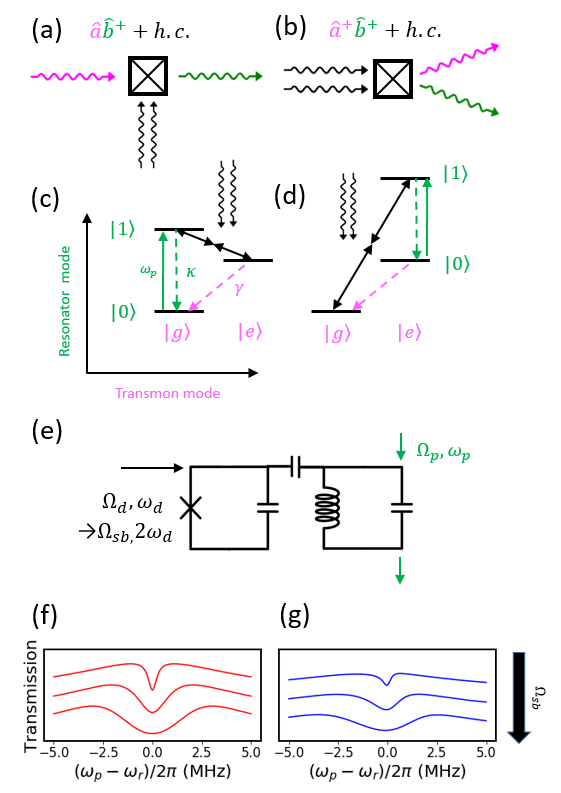}
    \caption{Overview of the experiment. (a,b) Schematics of the beam splitter and two-mode squeezing interactions. Black-, magenta- and green-wavy arrows indicate driving, transmon mode and resonator mode photons, respectively. (a) Beam-splitter interaction. A photon in the transmon mode is converted to the resonator mode by two-photon driving. (b) Two-mode squeezing interaction. Two-photon driving creates both transmon and resonator photons simultaneously. (c,d)  Energy level diagrams. Dashed-arrows indicate decay of transmon and resonator. The solid-green arrow represents probe tone through the resonator. (e) Simplified circuit diagram of the device. (f,g) Electromagnetically induced transparency (EIT) spectrum of the resonator calculated by the numerical model in \ref{eq3} when $\Omega_{sb}/2\pi$ is 2, 4, and 6 MHz, respectively (from top to bottom). Red- and blue-curves correspond to the BS and the TMS interaction respectively. See main text for detail simulation conditions}
    \label{fig:overview}
\end{figure}

\section{Experiment}
\label{3}

Both the BS and the TMS interaction are schematically described in Fig.~\ref{fig:overview}a and Fig.~\ref{fig:overview}b. Two black-wavy arrows indicate the two-photon drive. Fig.~\ref{fig:overview}c and Fig.~\ref{fig:overview}d denote energy diagram descriptions. In all of the descriptions, the resonator and the transmon mode are colored green and magenta, respectively. In addition to the two-photon drive, we have a weak probe field (green) through the resonator mode to estimate $\Omega_{sb}$ through the resonator's response.
The decay rates of both modes are $\kappa$ and $\gamma$, respectively. 
The energy levels of the resonator mode are denoted by $\ket{0}$, $\ket{1}$, $\ket{2}$,... and those of the transmon mode are denoted by $\ket{g}$, $\ket{e}$,...

Fig.~\ref{fig:overview}e depicts a simplified circuit diagram of the system. We drive the transmon mode through a direct driveline and we probe the resonator mode through another feedline coupled to the resonator. Fig.~\ref{fig:overview}f and Fig.~\ref{fig:overview}g show how the probe transmission through the resonator varies with increasing $\Omega_{sb}$ for both BS (f) and TMS (g) interaction. The curves are obtained by solving a numerical model based on Eq.~\ref{eq3} with dissipation operators. The decay rates of the resonator and transmon modes in the calculation are $\kappa/2\pi \approx 10.2$ MHz and $\gamma/2\pi \approx 129$ kHz. These parameters are similar to those in the experiment. $\Omega_{sb}/2\pi$ is set by 2, 4, and 6 MHz in both BS and TMS interaction.
The detailed information on the experimental setup and device is provided in Appendix~\ref{append-3}.

In the experiment, we deliberately design a large $\kappa$ to facilitate the detection of the interactions through the resonator's transmission, even with small $\Omega_{sb}$. Our system satisfies the condition for electromagnetically induced transparency (EIT) \cite{eit} as long as $\Omega_{sb}$ is smaller than $\left | \gamma-\kappa \right|$. In this regime, $\Omega_{sb}$ and the other parameters independently shape the transparency window in the middle of the transmission spectrum of the resonator. Thereby, we extract $\Omega_{sb}$ by fitting the resonator's transmission. The resonator's linewidth is overwhelmingly larger than the linewidth of the qubit and therefore the system is in the EIT condition as long as $\Omega_{sb}$ is less than around 10 MHz.

The observed $\omega_{t,r}$ are $2\pi\times$ 6.8112 and 4.0755 GHz, respectively. The observed $A_{t}$ is $2\pi\times 150$ MHz and $A_{r}$ can be deduced by $A_{tr}\approx\sqrt{A_{t}A_{r}}$.
Since the resonator has a broad linewidth, we cannot simply extract $A_{tr}$ from the photon number by splitting the resonator or the transmon spectrum. 
We obtained $A_{tr}/2\pi \approx$ 497 kHz from another calibration method in Appendix~\ref{append-4}. From these observations, we can calculate the system's parameters in Eq.~\ref{eq0} and Eq.~\ref{eq1}. The obtained values are $(\omega_{t}^{(0)},\omega_{r}^{(0)},\omega_{t}^{(1)},\omega_{r}^{(1)},g)=2\pi\times(6.8131,4.0823,6.81755,4.075953,0.1207)$ GHz and $(\chi_{t},\chi_{tr})=2\pi\times(137.4,0.384)$ MHz.

\begin{figure}
    \centering
    \includegraphics[width=1.1\linewidth]{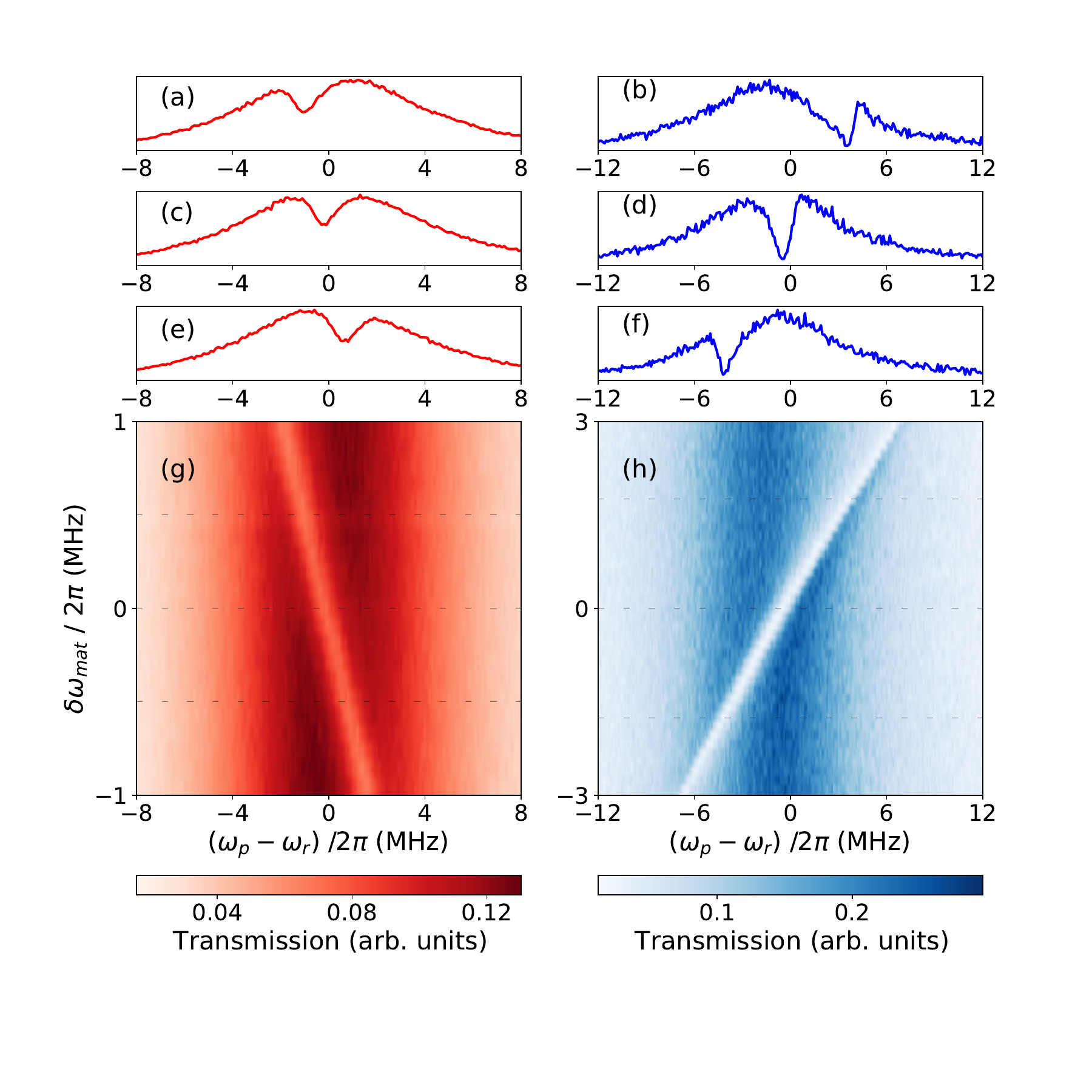}
    \caption{Spectroscopic observation of sideband transitions. (g) Drive frequency sweep around the matching condition for beam splitter (BS) interaction. (h) The same for two-mode squeezing (TMS) interaction. $\delta\omega_{mat}$ is 
    the deviation of the driving frequencies from the matching conditions ($\omega_d-\omega'_{mat}$).
    (a-f) The cross section of the dashed lines in (g) and (h), respectively. The y-axes refer to the resonator's transmission.}
    \label{fig:ResScan}
\end{figure}

\begin{figure}
    \includegraphics[width=0.5\textwidth]{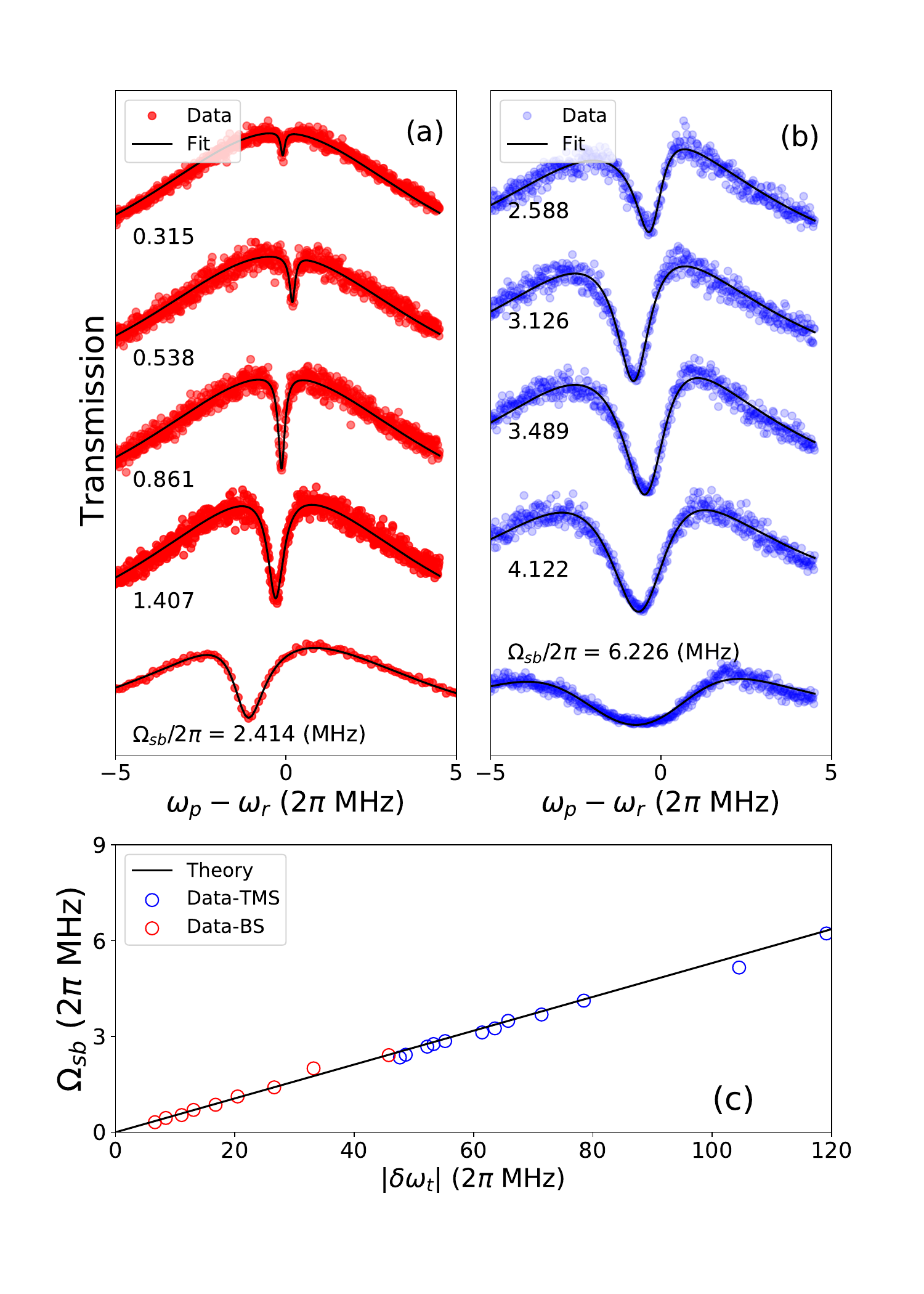}
    \caption{Driving power dependence of sideband transition rates.
    Left-hand (red) : BS interaction. Right-hand (blue) : TMS interaction.
    (a,b) Resonator transmission spectrum (circle) with increasing driving amplitude (top to bottom) while keeping $\omega'_{mat}\approx\omega_{d}$. Solid curves are fits to the data based on numerical model.
    From these fits, we extracted $\Omega_{sb}$ and $\delta\omega_{t}$.
    The probe amplitude $\Omega_p$ is 2$\pi\times$130.6 kHz, except that $\Omega_p$ is 10 times larger for the lowest dataset of (a). 
    (c) 
    Observed $\Omega_{sb}$ with respect to corresponding $\delta\omega_{t}$ (circles). The solid line indicates a theory based on Eq.~\ref{eq3-1}. Fitting errors in $\Omega_{sb}$ are around 1\% and are not plotted in the figures.
    }
    \label{fig:waterfall}
\end{figure}

In Fig.~\ref{fig:ResScan}, we present the procedure used for determining the frequency matching conditions. We define $\omega_{mat}$ that satisfies $2\omega_{mat}=|\omega_{t}\pm\omega_{r}|$ for both the BS and TMS interactions.
In reality, the resonances undergo shifts, $\omega_{t,r}\rightarrow\omega_{t,r}'= \omega_{t,r}+\delta\omega_{t,r}$ and in our system we have $\delta\omega_{t} \gg \delta\omega_{r} \approx 0$.
Thus, we have modified matching conditions, $2\omega'_{mat}=|\omega_{t}'\pm\omega_{r}|$.
We swept the driving frequency $\omega_{d}$ and find the condition $\omega_{d}\approx\omega'_{mat}$. We obtain the matching conditions when the transparency window is located at the center in the transmission spectrum. 
More quantitatively, $\omega'_{mat}$ can be obtained by extracting $\omega'_{t}$ when fitting the transmission data with numerical model given in Appendix~\ref{append-1-3}. 
Roughly, $\omega'_{mat}/2\pi\approx$ 1.36 and 5.44 GHz are expected for both the BS and the TMS interaction, respectively.
For the BS interaction, $\omega'_{mat}$ is extremely far off-resonant ($\Delta/\Sigma \approx$ 0.6). This regime of the driving parameter has not been explored. Meanwhile, for TMS interaction, $\omega'_{mat}$ is relatively closer to the RWA regime ($\Delta/\Sigma \approx$ 0.11).

In Fig.~\ref{fig:waterfall} a–b, we plot a portion of the transmission spectrum observed in the experiment. We scan the sideband driving power preserving the condition $\omega_{d}\approx\omega'_{mat}$.
The solid curves are the fits based on the numerical model that we used in Fig.~\ref{fig:overview}f–g. In the fitting process, the free parameters are $\Omega_{sb}$, $\gamma$ and $\delta\omega_{mat}$, while the other parameters are fixed. As we increase the driving amplitudes, we can readily see that the transparency windows behave as expected from Fig.~\ref{fig:overview}f–g.
In Fig.~\ref{fig:waterfall}c, we plot $\Omega_{sb}$ with respect to the corresponding $\delta\omega_{t}$, both of which are extracted from the fitting. The statistical errors in extracting $\Omega_{sb}$ from the fitting are around only 1\%, and thus not presented in the figures.
We can find a linear correlation between $\delta\omega_t$ and $\Omega_{sb}$ . The slope of the solid line is obtained from Eq.~\ref{eq3-1}, with no free parameters. 
It is of note that both BS and TMS data lie on the same theoretical plot, although the driving parameters for each lives in distinct regimes.

To directly identify the breakdown of the RWA, we need to calibrate $\Omega_d$ from an independent method not relying on the transmon frequency shifts. If we know the microwave power at the device ($P_{d}$), and the coupling rate between the transmon and drive line ($\gamma_{ex}$), then $\Omega_d$ is simply given by $\sqrt{P_{d}\gamma_{ex}/\hbar\omega_d}$. However, the uncertainty in the driveline attenuation sets a challenge. An error of only 1 dB in the attenuation induces a 10$\%$ error in $\Omega_d$, which is critical to our study. In  future research, this challenge can be circumvented by using an additional `sensor' qubit, as recently demonstrated in \cite{Zhou-PRAppl-2020}.

\section{Numerical simulation}
\label{4}

\begin{figure}
    \centering
    \includegraphics[width=0.5\textwidth]{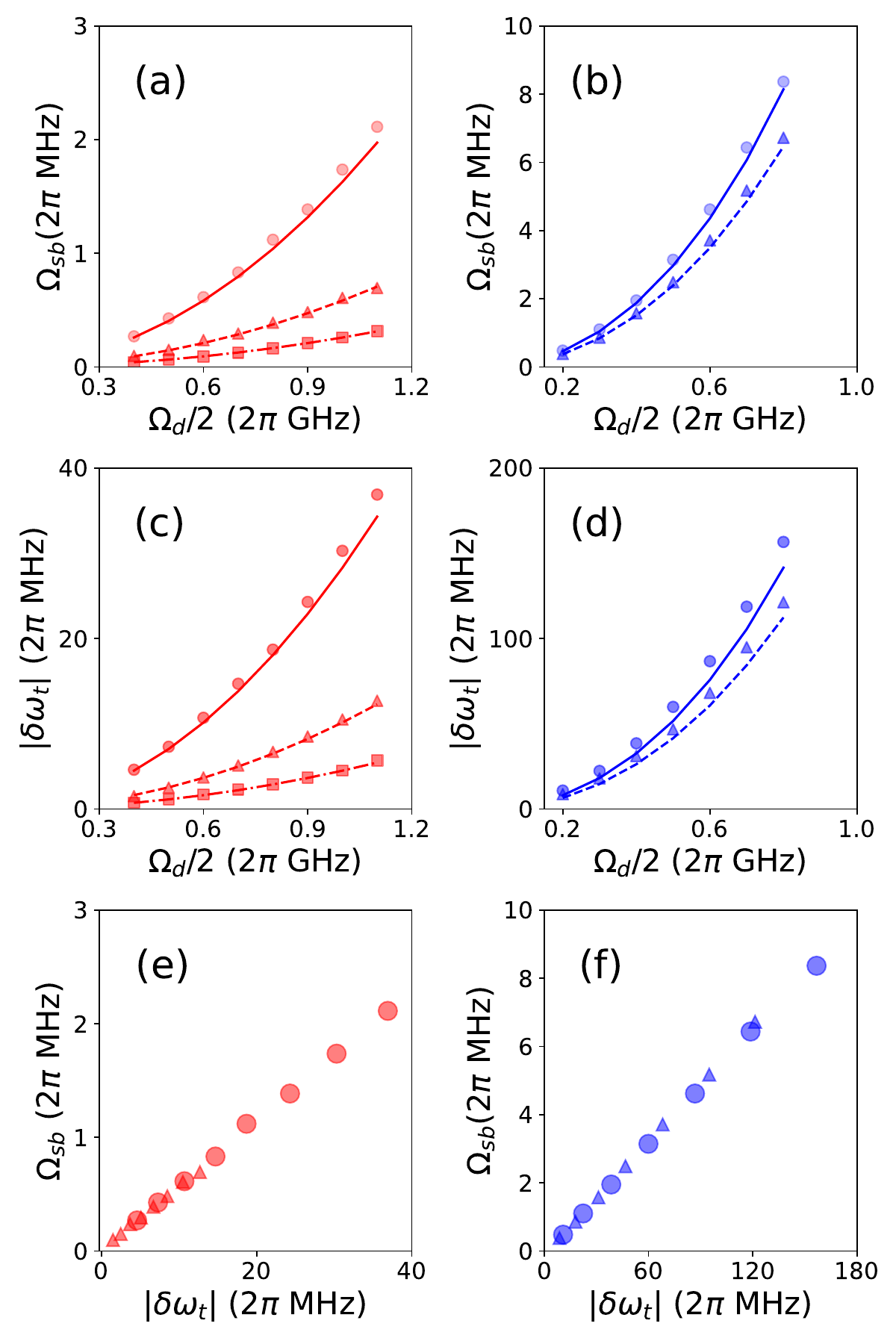}
    \caption{
    Influence of counter rotating terms of the driving on the sideband transition rates and the mode frequency shifts for BS (red, left-hand) and TMS (blue, right-hand) interactions.
    Symbols: numerical simulations with $\hat{H}_{tot}^{(0)}$ (circle), under the rotating wave approximation (RWA) keeping only co-rotating terms in $\hat{H}_{d}^{(0)}$ (triangle), and keeping only counter-rotating (CR) terms in $\hat{H}_{d}^{(0)}$ (squares). 
    Lines: analytical theory for $\hat{H}_{d}^{(0)}$ (solid), the RWA in $\hat{H}_{d}^{(0)}$ (dashed), and only CR terms in $\hat{H}_{d}^{(0)}$ (dotted).
    Dynamics beyond the RWA are clearly observed  in the transition rates (a,b) and frequency shift (c,d) for both red and blue sideband driving. 
    For the red sideband, even the purely counter-rotating terms lead to a non-zero coupling, although the deviation of the full result from the sum of the RWA and purley-CR calculations indicates additional contributions from correlations of the two.
    (e,f) Although discarding the CR terms leads to large correction of the $\Omega_{sb}$, the relationship between $\Omega_{sb}$ and $\delta \omega_t$ for the full $\hat{H}_{d}^{(0)}$ remains the same and falls on the same line (circles, triangles): the error that arises in discarding the CR terms is an incorrect value of both $\Omega_{sb}$ and $\delta \omega_t$ for a given and known driving strength.}
    \label{fig:experiment}
\end{figure}

We performed a comprehensive numerical analysis with the experimental conditions. We simulated the system's time domain dynamics by solving the $d\hat{\rho}/dt =-i[\hat{H}_{sys}^{(0)}+\hat{H}_{d}^{(0)},\hat{\rho}]$ without including any dissipation. $\hat{\rho}$ is the density matrix of the system. As in the experiment, we swept the driving frequency for a given $\Omega_d$ and find the frequency where a full oscillation takes place in transitions $\ket{e0}\Longleftrightarrow \ket{g1}$ (BS) or $\ket{g0} \Longleftrightarrow \ket{e1}$ (TMS). $\Omega_{sb}$ is then given by the frequency of the oscillation. More detailed descriptions on the method of the numerical simulation are given in Appendix~\ref{append-2}.  

In Fig.~\ref{fig:experiment}, we present the numerical calculation results (circles, triangles and squares) and corresponding analytical calculation results (solid, dashed, and dotted lines).
The circles and solid lines refer to the results with the $\hat{H}_{sys}^{(0)}+\hat{H}_{d}^{(0)}$.
In the plots, triangles and squares refer to the simulation results dropping the counter-rotating and co-rotating driving terms in $\hat{H}_d^{(0)}$. 
The analytical calculation is based on the Eq.~\ref{eq3-1}.
The dashed and dotted lines are obtained by Eq.~\ref{eq3-2} and Eq.~\ref{eq3-3}, respectively.

In Fig.~\ref{fig:experiment}a–b, we compare the sideband transition rates obtained by the numerical simulations (circles, triangles, and squares) with the analytically calculated values (solid, dashed, and dotted lines). 
In Fig.~\ref{fig:experiment} c–d, we present the frequency shifts of the transmon mode under the matching conditions for given the driving amplitudes in x-axes. The driving frequencies for each data points are set to satisfy the matching conditions for the given driving amplitudes. 
Although the RWA significantly distorts the $\Omega_{sb}$ and $\delta\omega_{t}$, the breakdown of the RWA is not visible in the $\Omega_{sb}$ versus $\delta\omega_{t}$ relation, as seen in Fig.~\ref{fig:experiment} e-f. The simulation data with the RWA perfectly lie on the data without the RWA. Therefore, a careful treatment is required when estimating $\Omega_{d}$ through $\Omega_{sb}$ or $\delta\omega_{t}$. Relying on the RWA results in significant overestimation of $\Omega_{d}$.

%Based on the numerically calculated values (filled circles), we extrapolate the relation between $\omega_{t}$ and $\Omega_d$ by polynomial fitting (solid lines). The analytically calculated $\delta\omega_{t}$ also well agree with the numerical values but small discrepancy exists at large driving amplitudes. 

\section{Conclusion}
In summary, we performed the quantitative investigation of two-photon assisted four-wave interactions in a superconducting circuit. Over the entire range of the driving amplitudes in this work, our theoretical, numerical and experimental values agree with each other, which suggests that the faithful quantitative estimation of sideband transition rates is possible.
This work expands our understanding in the strongly driven quantum systems. The findings through this work are not restricted to the system that we investigate here.

Kerr or Duffing type nonlinearity ubiquitously appears in many physics disciplines other than circuit QED, such as nonlinear optics, cavity optomechanics, and atomic physics \cite{Yurke-JOASB-1987,Bose-PRA-1997,Ludwig-PRL-2012,Gupta-PRL-2007}. Therefore, we believe our findings can influence a variety of types of research. This work also uses multi-photon assisted transition, which is widely adopted when a desired transition is dipole forbidden \cite{Eckhardt-PRL-1962,Hansch-PRL-1975,Eles-JCP-2004, Blais-PPA-2006,Wallraf-PRL-2007,Deppe-NPhys-2008,Leek-PRB-2009,Leek-PRL-2010,Kumar-Ncomm-2016,Premaratne-Ncomm-2017,Vepsalainen-SciAdv-2019,He-NPhys-2019,Meiling-JPCC-2019,Gasparinetti-PRL-2017,Gasparinetti-PRA-2020}. From the perspective that the studies of multi-photon transition beyond the RWA are mainly limited to theoretical cases \cite{Quattropani-PRA-1982,Saiko-JETP-2006, Meath-JCP-2018}, our work would attract attention.

\begin{acknowledgements}
We thank David Theron and Jochem Baselmans for providing us with NbTiN film. Byoung-moo Ann acknowledges support from the European Union’s Horizon 2020 research and innovation program under the Marie Sklodowska-Curie grant agreement No. 722923 (OMT).
This project also has received funding from the European Union’s Horizon 2020 research and innovation programme under grant agreement No. 828826 - Quromorphic.
The data that support the findings of this study are available in \cite{data}.
\end{acknowledgements}

\appendix
\section{Theoretical descriptions}
\label{append-theory}
\subsection{List of symbols}

Tab.~\ref{tab1} lists and defines the symbols that are used in the main text and supplemental material.

\begin{center}
\begin{table*}
 \begin{tabular}{||c c||} 
 \hline
 Symbol & Meaning \\ [0.5ex] 
 \hline\hline
 $\omega_{t,r}^{(0)}$ & Mode frequencies in uncoupled mode basis (Eq. (1) in the main text). \\ 
 \hline
 $\omega_{t,r}^{(1)}$ &  Mode frequencies in normal mode basis (Eq. (2) in the main text).\\
 \hline
 $\omega_{t,r}$ & Experimentally observed mode frequencies (Eq. (5) in the main text). \\
 \hline
 $\omega_{d}$ & Driving frequency. \\
 \hline
 $\Omega_{d}$ & Driving amplitude. \\
 \hline
 $\omega_{p}$ & Probe frequency. \\
 \hline
 $\Omega_{p}$ & Probe amplitude. \\
 \hline
 $\chi_{t,r}$ & Duffing nonlinearities of each modes.  \\
 \hline
 $A_{t,r}$ &  Anharmonicities of each modes.  \\
 \hline
 $A_{tr}$ & Cross-anharmonicity ($\approx\sqrt{A_{t}A_{r}}$). \\  
 \hline
 $\hat{\alpha}$ &  Transmon mode annihilation operator in uncoupled mode basis. \\
 \hline
 $\hat{\beta}$ & Resonator mode annihilation operator in uncoupled mode basis. \\
 \hline
 $\hat{a}$ & Transmon mode annihilation operator in normal mode basis.\\
 \hline
 $\hat{b}$ & Resonator mode annihilation operator in normal mode basis.\\ 
 \hline
 $\delta\omega_{t,r}$ & Frequency shifts of both modes under driving.\\ 
 \hline
 $\delta\omega_{t,r}$ & Frequency shifts of both modes under driving.\\ 
 \hline
 $\omega_{mat}$ & Sideband transition matching frequency without considering frequency shifts of modes.\\ 
 \hline
 $\omega'_{mat}$ & Sideband transition matching considering frequency shifts of modes.\\ 
 \hline
 $\delta\omega_{mat}$ & $\omega'_{mat}$ - $\omega_{d}$. \\
 \hline
 $\Omega_{sb}$ & Sideband transition rate.\\ 
 \hline
  $\gamma$ & Decay rate of the transmon mode.\\ 
 \hline
  $\kappa$ & Decay rate of the resonator mode.\\ 
 \hline
\end{tabular}
\caption{\label{tab1} Symbols and their definitions used in this work.}
\end{table*}
\end{center}

\subsection{Schrieffer–Wolff transformation}
We perturbatively diagonalize the total Hamiltonian by
applying the unitary transformation $\hat{U}(t)$ \cite{SW} to the total Hamiltonian $\hat{H}_{tot}^{(1)}=\hat{H}^{(1)}_{sys}+\hat{H}_{d}^{(1)}$, where $\hat{H}^{(1)}_{sys}$ and $\hat{H}_{d}$ are defined in the main text.
The transformed Hamiltonian $\hat{{H'}}_{tot}$ is given by,
\begin{equation}
\begin{split}
    \hat{{H'}}_{tot}^{(1)} = {\hat{U}\hat{H}\hat{U}^\dagger}+i(\partial_{t}\hat{U})\hat{U}^\dagger.
\end{split}
\end{equation}
Here, $\hat{U}(t)=e^{\xi(t)\hat{a}^\dagger-\xi(t)^{*}\hat{a}}$ 
and $\xi(t)=\frac{\Omega_{d}}{2\Delta}e^{-i\omega_{d}t}+\frac{\Omega_{d}}{2\Sigma}e^{i\omega_{d}t}$. $\hat{U}(t)$ simply displaces the field operator $\hat{a}$($\hat{a}^\dagger$) by -$\xi$(-$\xi^{*}$). Finally, $\hat{{H'}}_{tot}^{(1)}$ can be expressed by,
\begin{equation}
\begin{split}
    \hat{H'}_{tot}^{(1)} \approx ~ \,& ({\omega}_{t}^{(1)}+\chi_{t}) \hat{a}^\dagger \hat{a}+ {\omega}_{r}^{(1)} \hat{b}^\dagger \hat{b}  \\
    & - \frac{1}{12} \left[ \chi_t^{1/4} (\hat{a} + \hat{a}^\dagger - \xi(t) -\xi^{*}(t)) + \chi_r^{1/4} (\hat{b} + \hat{b}^\dagger) \right] ^4.
\label{transformed}
\end{split}
\end{equation}
For given $\omega_d$, collecting the non-rotating terms at the transmon and resonator rotating frame in Eq.~\ref{transformed} yields Tab.~\ref{tab2}. We only list the terms that represent the interactions between different modes or the frequency shifts of each mode. 

% \subsection{Basis renormalization}
% $\hat{H}_{sys}^{(1)}$ in Eq.~\ref{eq1} contains many off-diagonal elements, and therefore the  

% in a subspace spanned by {$\ket{g0}, \ket{g1}, \ket{g2}, \ket{e0}, \ket{e1}, \ket{f0}$}. Here each $\ket{g,e,f}$ and $\ket{0,1,2}$ are the first, second, and third eigenstates of $\hat{a}^\dagger\hat{a}$ and $\hat{b}^\dagger\hat{b}$ respectively. We can define the eigenstates of $\hat{H}_{sys}^{(1)}$ confined in the subspace by {$\ket{g0}', \ket{g1}', \ket{g2}', \ket{e0}', \ket{e1}', \ket{f0}'$}.

\begin{center}
\begin{table*}[]
\begin{tabular}{|l|l|l|}
\hline
Operator (+ h.c)                                & Magnitude ($\times \Omega_{d}^2/4)$                                                                                                                                 & \begin{tabular}[c]{@{}l@{}}Matching condition\\ ($\omega_d\sim$)\end{tabular} \\ \hline
$\hat{a}^{\dagger}\hat{a}$              & \begin{tabular}[c]{@{}l@{}}$\chi_{t,r}$ \\ $\times (2\frac{1}{{\Delta}^2}+\frac{2}{\Delta\Sigma}+\frac{1}{{\Sigma}^2})$\end{tabular} & None                                                                          \\ \hline
$\hat{b}^{\dagger}\hat{b}$              & \begin{tabular}[c]{@{}l@{}}$\chi_{t,r}$ $\times (2\frac{1}{{\Delta}^2}+\frac{2}{\Delta\Sigma}+\frac{1}{{\Sigma}^2})$\end{tabular} & None                                                                          \\ \hline
$\hat{a}\hat{b}^{\dagger}$      & $\chi_{t}^{3/4}\chi_{r}^{1/4}$ $\times (\frac{1}{{\Delta}^2} +\frac{2}{\Delta\Sigma}+\frac{1}{{\Sigma}^{2}})$                      & $|\omega_{t}^{(1)}-\omega_{r}^{(1)}|/2$                                                   \\ \hline
$\hat{a}^{\dagger}\hat{b}^{\dagger}$    & $\chi_{t}^{3/4}\chi_{r}^{1/4}$ $\times (\frac{1}{{\Delta}^2} +\frac{2}{\Delta\Sigma}+\frac{1}{{\Sigma}^{2}})$                      & $\omega_{t}^{(1)}+\omega_{r}^{(1)}/2$                                                   \\ \hline
$\hat{a}\hat{b}^{\dagger 2}$            & $\chi_{t}^{1/4}\chi_{r}^{3/4}$  $\times (\frac{1}{\Delta} +\frac{1}{\Sigma})$                                                      & $|2\omega_{r}^{(1)}-\omega_{t}^{(1)}|$                                                    \\ \hline
$\hat{a}^{\dagger}\hat{b}^{\dagger 2} $ & $\chi_{t}^{1/4}\chi_{r}^{3/4}$ $\times (\frac{1}{\Delta} +\frac{1}{\Sigma})$                                                      & $2\omega_{r}^{(1)}+\omega_{t}^{(1)}$                                                      \\ \hline
$\hat{a}^{2}\hat{b}^{\dagger}$          & $\chi_{t}^{3/4}\chi_{r}^{1/4}$ $\times (\frac{1}{{\Delta}} +\frac{1}{{\Sigma}})$                                                   & $|2\omega_{t}^{(1)}-\omega_{r}^{(1)}|$                                                    \\ \hline
$\hat{a}^{\dagger 2}\hat{b}^{\dagger} $ & $\chi_{t}^{3/4}\chi_{r}^{1/4}$ $\times (\frac{1}{{\Delta}} +\frac{1}{{\Sigma}})$                                                   & $2\omega_{t}^{(1)}+\omega_{r}^{(1)}$                                                    \\ \hline
\end{tabular}
\caption{\label{tab2} List of a portion of the non-rotating terms at the transmon and resonator rotating frame for given $\omega_d$ derived from the fourth power term of Eq.~\ref{transformed}.}
\end{table*}
\end{center}

\subsection{Modeling transmission spectrum}
\label{append-1-3}
The resonator transmission spectrum is proportional to $\textup{Tr}_{t}[\hat{\rho}_{ss}\hat{b}]$.
Here, $\hat{\rho}_{ss}$ is a steady state density matrix of the transmon and resonator system, and $\textup{Tr}_{t}$ indicates trace over the transmon states. $\hat{\rho}_{ss}$ can be calculated based on the below Eq.~\ref{model},
\begin{equation}
\begin{split}
   	\frac{d\hat{\rho}}{dt} =  \vphantom{\sum_n} -\frac{i}{\hbar} \left[ \hat{H}_{\textup{low}}+\hat{H}_{p}(t), \hat{\rho}(t) \right]\\
	+\frac{\gamma}{2}\mathcal{D}[\hat{a}]\hat{\rho} +  \frac{\kappa}{2}\mathcal{D}[\hat{b}]\hat{\rho}.
\end{split}
\label{model}
\end{equation}
$\hat{H}_{p}(t)=\Omega_p\cos{(\omega_p t)}$ is the Hamiltonian of the prove field. 
$\mathcal{D}[\mathcal{\hat{O}}]\hat{\rho}$ is defined by $2\mathcal{\hat{O}}\hat{\rho}\mathcal{\hat{O}}^\dagger - \mathcal{\hat{O}}^\dagger\mathcal{\hat{O}}\hat{\rho}-\hat{\rho}\mathcal{\hat{O}}^\dagger\mathcal{\hat{O}}$.
$\kappa$ is the decay rate of the resonator mode, and $\gamma$ is that of the transmon mode. 
We neglect the pure dephasing rate of the transmon mode. Since we employ a single Josephson junction design, it is expected that the coherence time of the transmon mode is only limited to the decay time.
For a steady state, we have $\frac{d\hat{\rho}_{ss}}{dt}=0$, then we can calculate $\hat{\rho}_{ss}$ from Eq.~\ref{model}. 

Transmission spectrum is a function of a set of variables ($\omega_d$, $\omega_p$, $\Omega_p$, $\Omega_{sb}$, $\omega'_{t}$, $\omega_{r}$, $A_t$, $A_r$, $A_{tr}$ $\kappa$, and $\gamma$). Here, $\omega_p$ is the independent variable in the fitting process. We fix $\kappa$, $\omega_{r}$, $A_t$, $A_r$, and $A_{tr}$ by the values we obtain from the independent measurement without driving field. These quantities are hardly shifted under the driving. $\omega_d$ is given by the experiment. The free fitting parameters are $\omega_{t}$, $\Omega_{sb}$, $\Omega_p$, and $\gamma$. These quantities are extracted from the fitting process.

\section{Numerical simulations}
\label{append-2}
\begin{figure}
    \centering
    \includegraphics[width=0.8\linewidth]{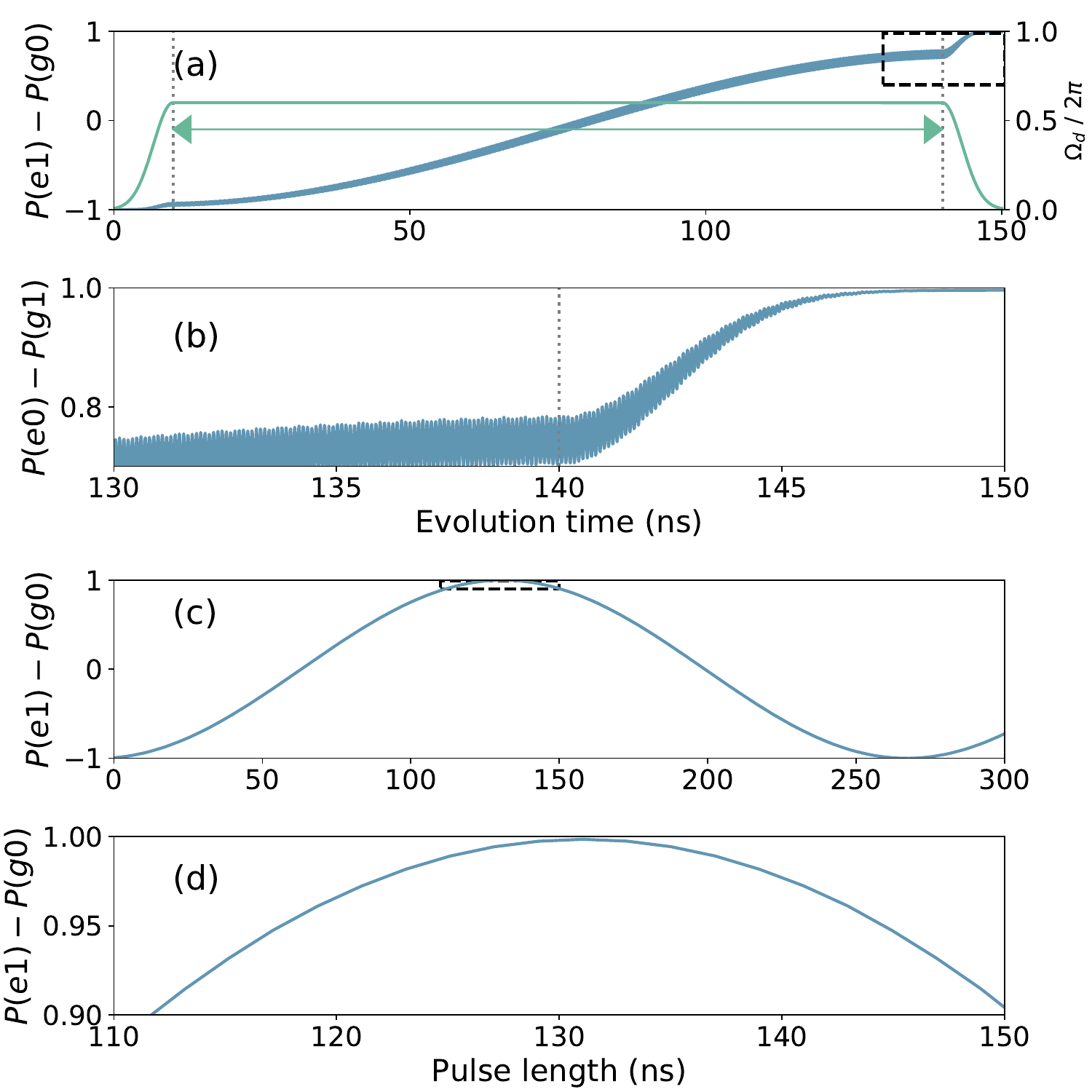}
    \caption{Illustration of the method of the numerical simulations. (a) The blue line indicates the dynamics of the system when the driving frequency satisfies the matching condition of the TMS interaction for given driving amplitude ($2\pi\times 300$ MHz). $P(e1)$ and $P(g0)$ refer to the population of each state. We assume the time-dependence in the sideband driving amplitude (green line), with 10 ns rising and falling time. The definition of the driving pulse length is graphically depicted by a green arrow. We adjust the pulse length such that almost a full state transfer from $\ket{e1}$ to $\ket{g0}$ takes place. (b) The area enclosed by the dashed square in (a) is zoomed in. A significant change in $P(e1)$-$P(g0)$ can be identified. (c) We repeat the simulation with a various pulse length and plot $P(e1)$-$P(g0)$ at the end of each pulse. (d) The area enclosed by the dashed square in (c) is zoomed in. The fidelity of the state transfer from $\ket{e1}$ to $\ket{g0}$ is 99.85 \%.}
    \label{fig:convergent}
\end{figure}

\begin{figure}
    \centering
    \includegraphics[width=1\linewidth]{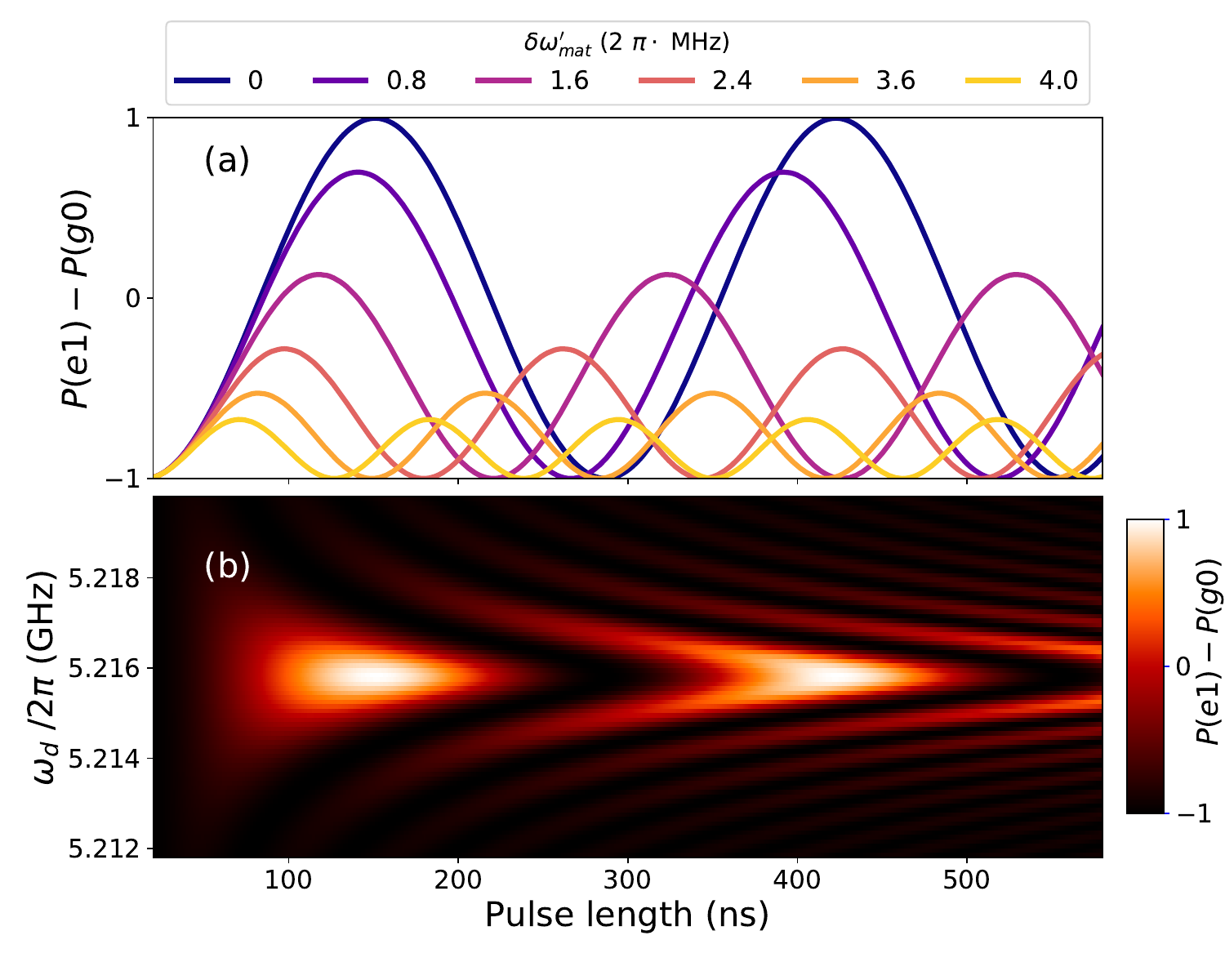}
    \caption{Driving frequency sweep in the numerical simulation. (a) Time-domain simulations of TMS interaction with various $\delta\omega'_{mat}$. (b) Continuously scanning the driving frequency near the matching condition. Given driving amplitude is $2\pi\times 300$ MHz. See the text for the system parameters that are used in the simulation.}
    \label{fig:fsweep}
\end{figure}

\begin{figure}
    \centering
    \includegraphics[width=1\linewidth]{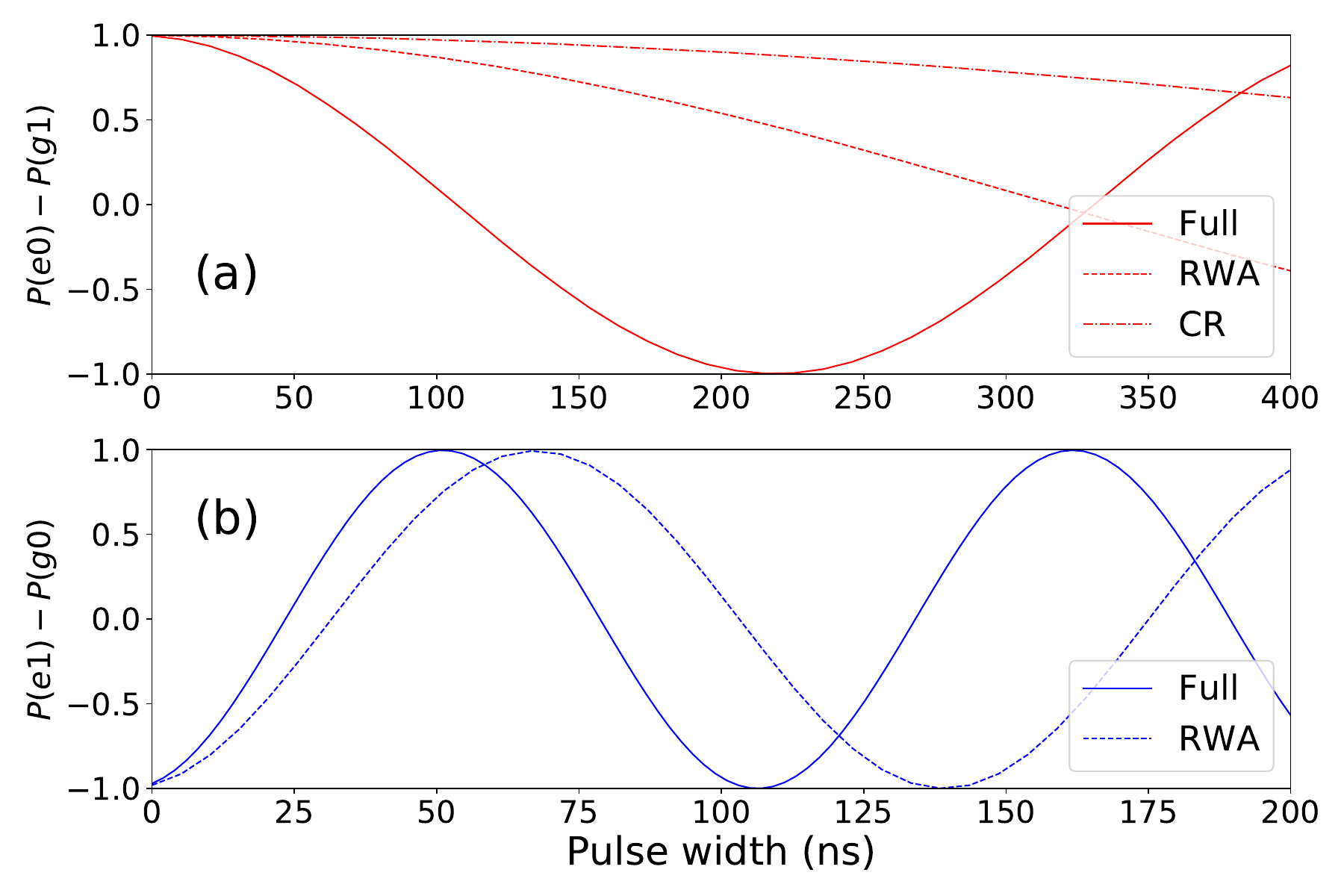}
    \caption{Time-domain plots of the numerical simulation results. The plots in Top (a) and bottom (b) panels are time-domain dynamics of the BS and TMS interactions respectively.
    In both cases, $\omega_q^{(0)} = 2\pi \times$ 6.5 GHz $\omega_q^{(0)} = 2\pi \times$ 4.0 GHz, $g/2\pi=$200 MHz, $\chi_{t}/2\pi$=200 MHz and $\Omega_{d}/2\pi$=600 MHz are chosen.
    Full : Simulations with $\hat{H}_{tot}^{(0)}$. RWA : With only co-rotating driving terms in $\hat{H}_{d}^{(0)}$, CR : With only counter-rotating driving terms in $\hat{H}_{d}^{(0)}$}.
    \label{fig:extendsim-TD}
\end{figure}

In this section, we describe the detail procedures of the time-domain numerical simulations. The dynamics of the system are governed by the equation, $d\hat{\rho}_{sys}/dt = -i[\hat{H}_{sys}^{(0)}+\hat{H}_{d}^{(0)}(t),\hat{\rho}_{sys}]$, where $\hat{H}_{sys}^{(0)}$ and $\hat{H}_{d}^{(0)}$ follow the same definition in the main text. Here, $\hat{\rho}_{sys}$ is density matrix of the transmon and resonator. We do not take the dissipation into consideration in the time-domain dynamic simulations.
Fig.~\ref{fig:convergent}a shows the simulated dynamics (blue line) when the driving frequency satisfies the matching condition for two-mode squeezing (TMS) interaction. The system parameters used in the simulation are the same with the experimental conditions. The sideband drive ($\Omega_{d}(t)$, green line) is given as a pulse with 10-ns of Gaussian rising and falling. The arrow indicates the length of the pulse.
Fig.~\ref{fig:convergent}b shows the area enclosed by the dashed  square in Fig.~\ref{fig:convergent}a. One can identify the qubit and resonator states significantly vary during the rising and falling duration of the sideband pulse. 
In Fig.~\ref{fig:convergent}c, we sweep the length of the sideband pulse and plot the states of the system at the end of the pulse. We obtain a clear sinusoidal curve. Fig.~\ref{fig:convergent}d shows the area enclosed by the dashed  square in Fig.~\ref{fig:convergent}c.  

We sweep the driving frequency for each simulation data point and find the optimal frequency that yields the resonant sideband transitions. This procedure is described in Fig.~\ref{fig:fsweep}. We chose the $w_d$ when the oscillation has a maximum contrast.
We present the simulation data with different driving Hamiltonian in Fig.~\ref{fig:extendsim-TD}. 
The solid lines refer to the results with a full driving Hamiltonian containing both co- and counter-rotating terms. 
The dotted lines (dashed lines) are obtained by the simulations with only co-rotating (counter-rotating) terms in the driving Hamiltonian. See the caption for the detail conditions in the simulations.

\section{Experimental methods}
\subsection{Experimental setup}
\label{append-3}
\begin{figure}
    \centering
    \includegraphics[width=0.8\linewidth]{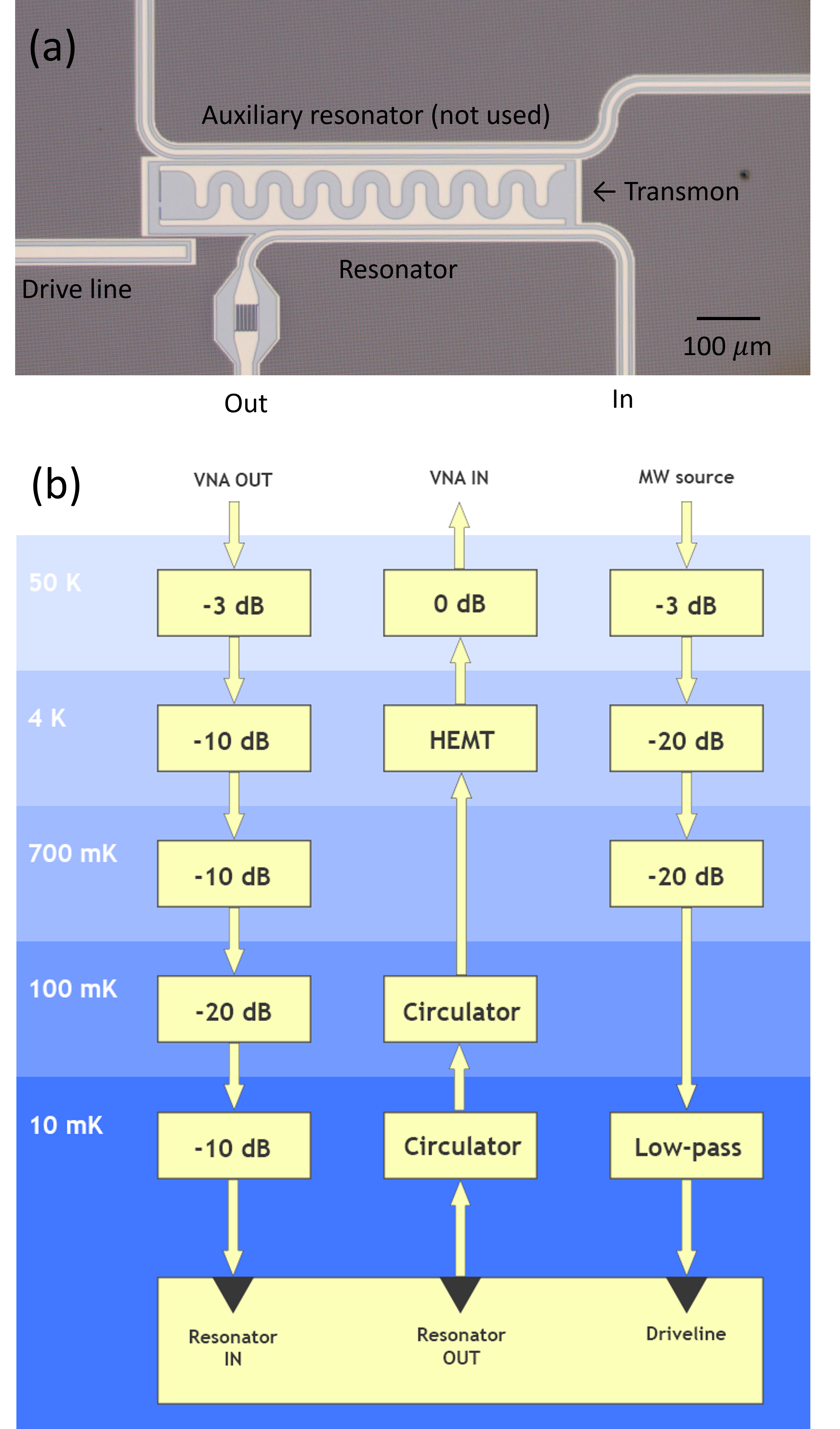}
    \caption{Experimental setup. (a) An optical microscope image of the device used in the experiment. (b) A cryogenic wiring diagram and measurement electronics.}
    \label{fig:setup}
\end{figure}

An optical microscope image of the device is given in Fig.~\ref{fig:setup}a. The device is comprised of a transmon and two co-planar waveguide resonators. The design of the device is the same as the one used in our previous work \cite{bann-2020}. Only one of the resonators was used in this experiment. In addition, there is a drive line directly coupled to the transmon. 
The base layer of the circuit is fabricated from 100 nm niobium titanium nitride (NbTiN) film on a Silicon substrate.
The detailed procedure to prepare the NbTiN film is described in \cite{SRON}.
The transmon is comprised of a Al-AlOx-Al Josephson junction and a finger capacitor. The transmon is not flux tuneable and therefore the frequency is insensitive to the external magnetic field noise. 

A cryogenic wiring diagram and measurement electronics are given in Fig.~\ref{fig:setup}b.
The device is mounted at the mixing chamber plate of a Bluefors LD-400 dilution fridge. The temperature of the  plate is around 10 mK during the measurements. The device is enclosed within a cylindrical cooper shield to block the infrared radiation. To block the external magnetic fields, the copper can is enclosed by a Aluminum shield and two Mu-metal shields. The shields are not represented in the figure.
We used a vector network analyzer (Keysight N5222A) to measure the resonator transmissions. An additional microwave source (Keysight N5183B) was used for sideband drivings. We used a  non-dissipative low pass filter (Minicircuit VFL-3800+) in the drive line (third column).

\subsection{Device parameter extraction}
\label{append-4}
\begin{figure}
    \centering
    \includegraphics[width=0.9\linewidth]{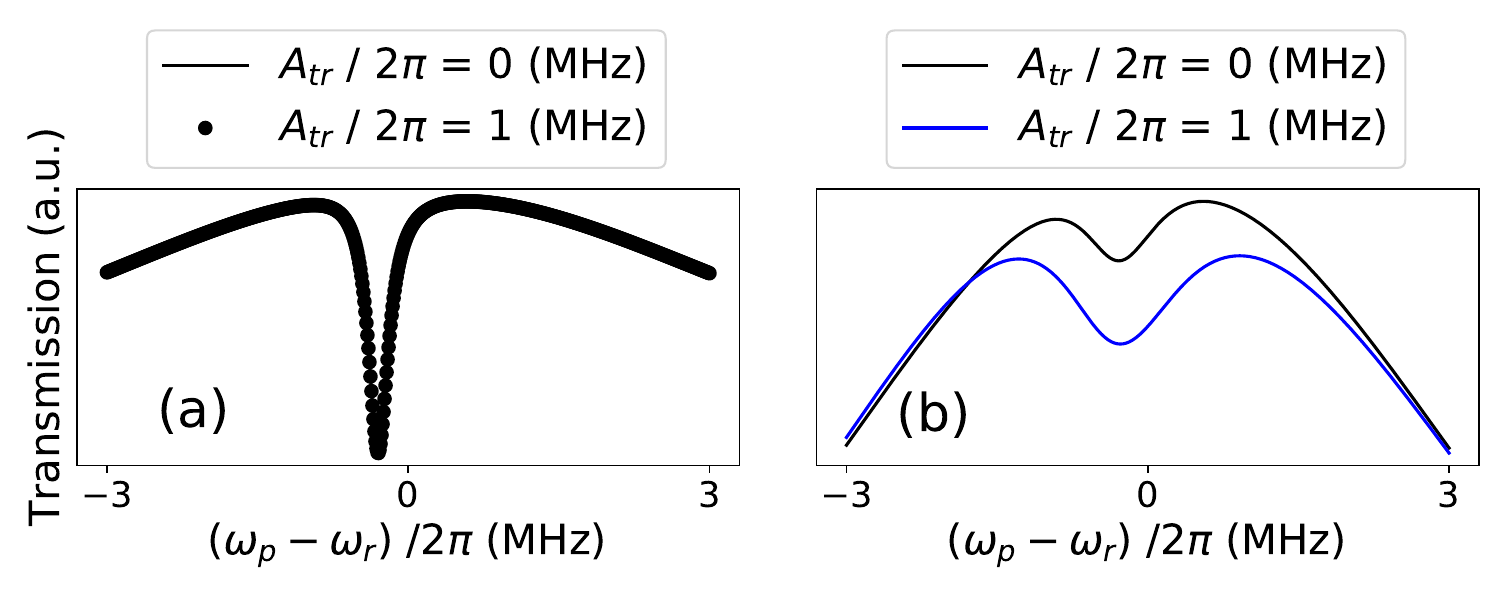}
    \caption{Effect of the cross-anharmonicity ($A_{tr}$) in the resonator transmission when a beam splitter interaction is applied. The beam splitter interaction between the transmon and resonator modes are applied in the simulation. (a) In the linear response regime (weak probe, $\Omega_p/2\pi=10$kHz), the cross-anharmonicity does not make a difference in the spectrum. (b) In the nonlinear regime (strong probe, $\Omega_p/2\pi=3$MHz), we can easily confirm the effect of the $A_{tr}$ from the spectrum.}
    \label{fig:cal1}
\end{figure}

\begin{figure}
    \centering
    \includegraphics[width=1.1\linewidth]{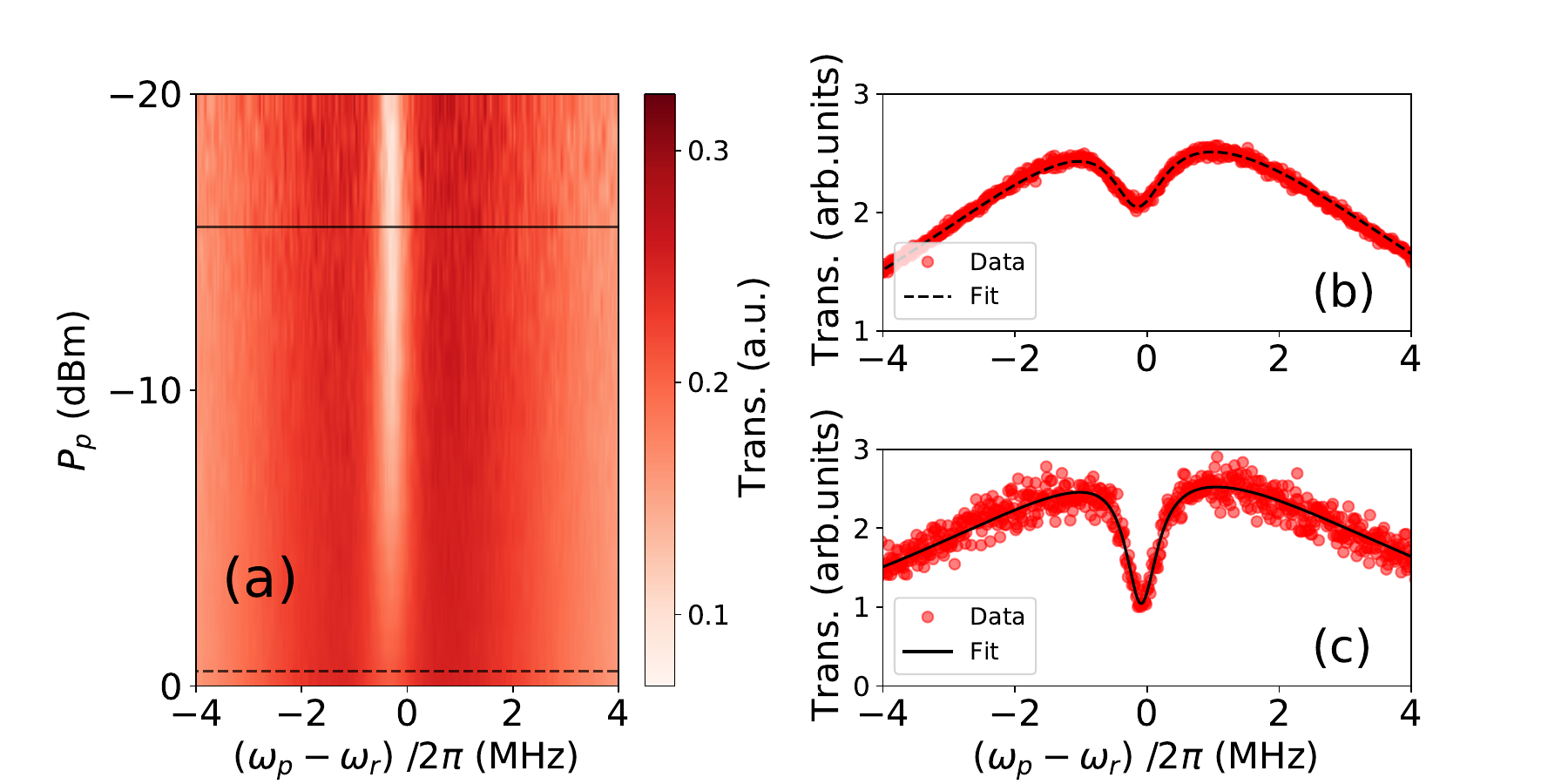}
    \caption{Calibration of the cross-anharmonicity ($A_{tr}$). (a) The transmission spectrum of the resonator while scanning probe power ($P_{in}$). The transmon and the resonator modes are coupled by a beam splitter interaction. The horizontal dashed and solid lines indicate the data when $P_{in}$ is 0 dBm and -15dBm, respectively. (b-c) The spectrum at the probe powers indicated in (a) with horizontal lines. The circles are experimental data and the solid curves are fits based on Eq.3 in the main text. From the linear response date (c), we extract $\Omega_{sb}$, $\omega_{d}$, $\omega_{r}$,$\kappa$ and $\gamma$ by the fitting. When fitting the data in (b), these quantities are fixed with the extracted values obtained from (c). Then we extract $\Omega_{p}$ and $A_{tr}$}.
    \label{fig:cal2}
\end{figure}

\begin{figure}
    \centering
    \includegraphics[width=0.9\linewidth]{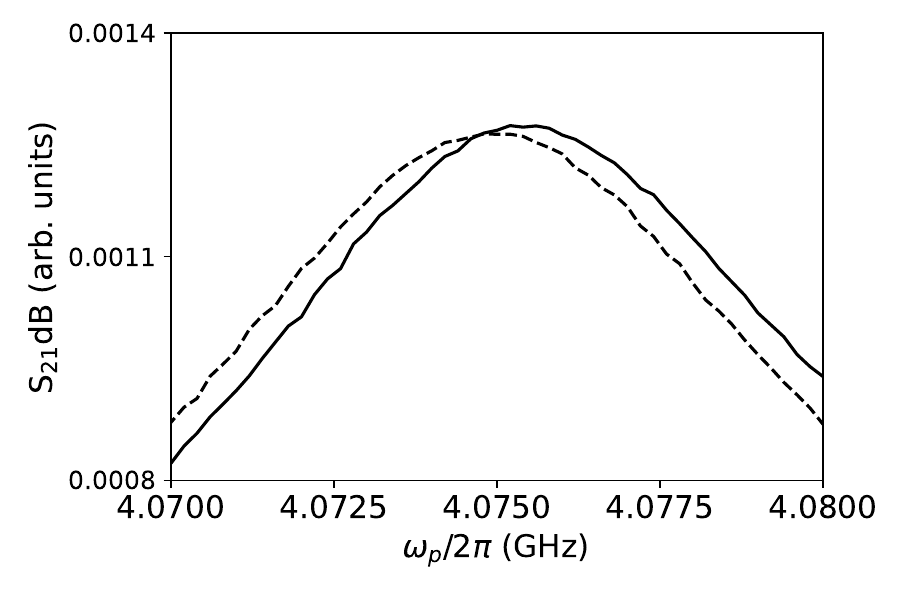}
    \caption{Calibration of the cross-anharmonicity ($A_{tr}$) from the resonator's response. The lines indicate the resonator's transmission when the transmon is the ground state (solid) and approximately 50:50 mixed state between the ground and first excitation states (dashed). The resonance is shifted by 520 kHz.}
    \label{fig:cal3}
\end{figure}

In this section, we provide the procedure to calibrate the cross-anharmonicity ($A_{tr}$) between the transmon and resonator modes in the experiment. We use the fact that the EIT transmission spectrum of the resonator depends on the $A_{tr}$ in the nonlinear response regime.
In Fig.~\ref{fig:cal1}, we simulate the resonator's transmission spectrum with a beam splitter interaction ($\Omega_{sb}/2\pi$ 1.2 MHz). 
The model that we used in the simulation is based on Eq. (3) in the main text including dissipation operators.
In addition, we set $\delta\omega_{mat}/2\pi$ by -300 kHz. In the simulation, the linewidths of the resonator and transmon modes are the same with those in the experiment.
We simulate in both linear response (Fig.~\ref{fig:cal1}a) and nonlinear response (Fig.~\ref{fig:cal1}b) regimes. In the linear response regime, we cannot distinguish the $A_{tr}$ from the transmission. Meanwhile, the effect of the $A_{tr}$ is prominent in the nonlinear response regime.

Fig.~\ref{fig:cal2}a shows the measured resonator transmission spectrum while sweeping the probe power. $P_{p}$ is the resonator probe power measured at the output port of the vector network analyzer (VNA). 
Note that the contrast of the transparency window near the center decreases with increasing probe power. We first fit the resonator's transmission data in the linear response regime (solid line), setting $\Omega_{sb}$, $\omega_{d}$, $\omega'_{t}$, $\kappa$ and $\gamma$ as free parameters.  
Then, we fit the data in the nonlinear response regime (dashed line) while fixing all the parameters obtained from the first fitting and only $\Omega_{p}$ and $A_{tr}$ are free fitting parameters.
When fitting the data in the linear response regime, we set $A_{tr}=0$ and $\Omega_{p}/2\pi$ = 10 kHz. The choice of $A_{tr}$ can be justified since we already know $A_{tr}$ hardly affects the transmission in the linear response regime.
The fitting results in both regimes are given in Fig.~\ref{fig:cal2}b and Fig.~\ref{fig:cal2}c.
We obtain $A_{tr}/2\pi$ = 497 kHz and $\Omega_{p}/2\pi$ = 4.35 MHz from the data in the nonlinear regime.
 
We can also obtain $A_{tr}$ from the fact that the resonator's transition frequency depends on the transmon's quantum states \cite{Blais-PPA-2006}. Fig.~\ref{fig:cal3} shows how the resonator's transmission spectrum changes as we populate the transmon's first excited state. We drive the transmon mode with its resonant frequency and increase the power until we cannot see any further shift in the resonator's frequency. With this drive power, we can approximate the transmon's state 50:50 mixed state between the ground and first excited states. We observe a frequency shift of 520 kHz, which can be interpreted as $A_{tr}$.

$A_{tr}$ extracted from Fig.~\ref{fig:cal3} is slightly larger than the value obtained from Fig.~\ref{fig:cal2}. The discrepancy of the expected sideband transition rates based on both is about 2 percent.
In the main text, we use $A_{tr}/2\pi$ =  kHz obtained from Fig.~\ref{fig:cal2}. This approach is advantageous because we can extract the resonator probe power and $A_{tr}$ simultaneously, and consequently it guarantees more consistency.

\subsection{Transmon decay rate analysis}
\begin{figure}
    \centering
    \includegraphics[width=0.9\linewidth]{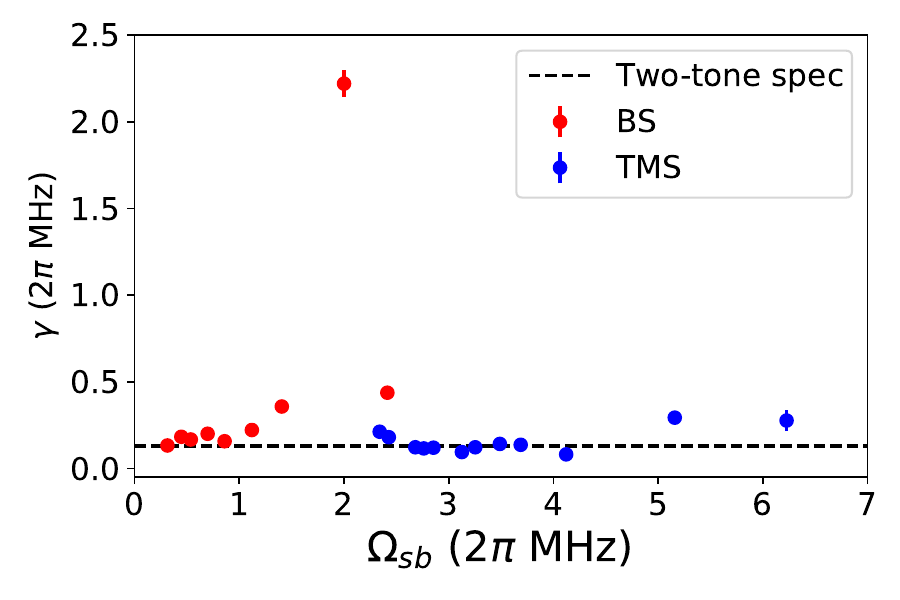}
    \caption{Comparison between the qubit decay rates ($\gamma$) extracted from the EIT spectrum fitting (dots) and the two-tone spectroscopy with a low probe and spectroscopy power (solid line).}
    \label{fig:gamma}
\end{figure}

In the fitting process to extract the sideband transition rates, the free fitting parameters other than $\Omega_{sb}$ are $\gamma$ and $\delta\omega_{mat}$. We also present the extracted values for $\gamma$ and $\delta\omega_{mat}$ in \cite{data}. In this section, we especially focus on the $\gamma$. Fig.~\ref{fig:gamma} shows the fitted $\gamma$ (dots) with respect to corresponding $\Omega_{sb}$. These values are consistent with the $\gamma$ from the low power two-tone spectroscopy (dashed line) in general. For BS interaction case, some data points far deviate from the dashed line. We attribute this to the undesired higher order sideband interactions. The matching frequency for BS interaction is close to the matching frequency for single-photon assisted sideband interaction between $\ket{e0}$ and $\ket{g2}$. Since the resonator mode has a much larger decay rate, this undesired interaction can increase the effective decay rate of the transmon mode. The rightmost two data of TMS interaction case also far deviate from the solid line. We cannot find the systematic reason for the discrepancy. We could attribute this to the fluctuation of the transmon's decay rate with respect to time.

\section{Additional analytical and numerical analysis}
\label{append-5}
\begin{figure}
    \centering
    \includegraphics[width=1\linewidth]{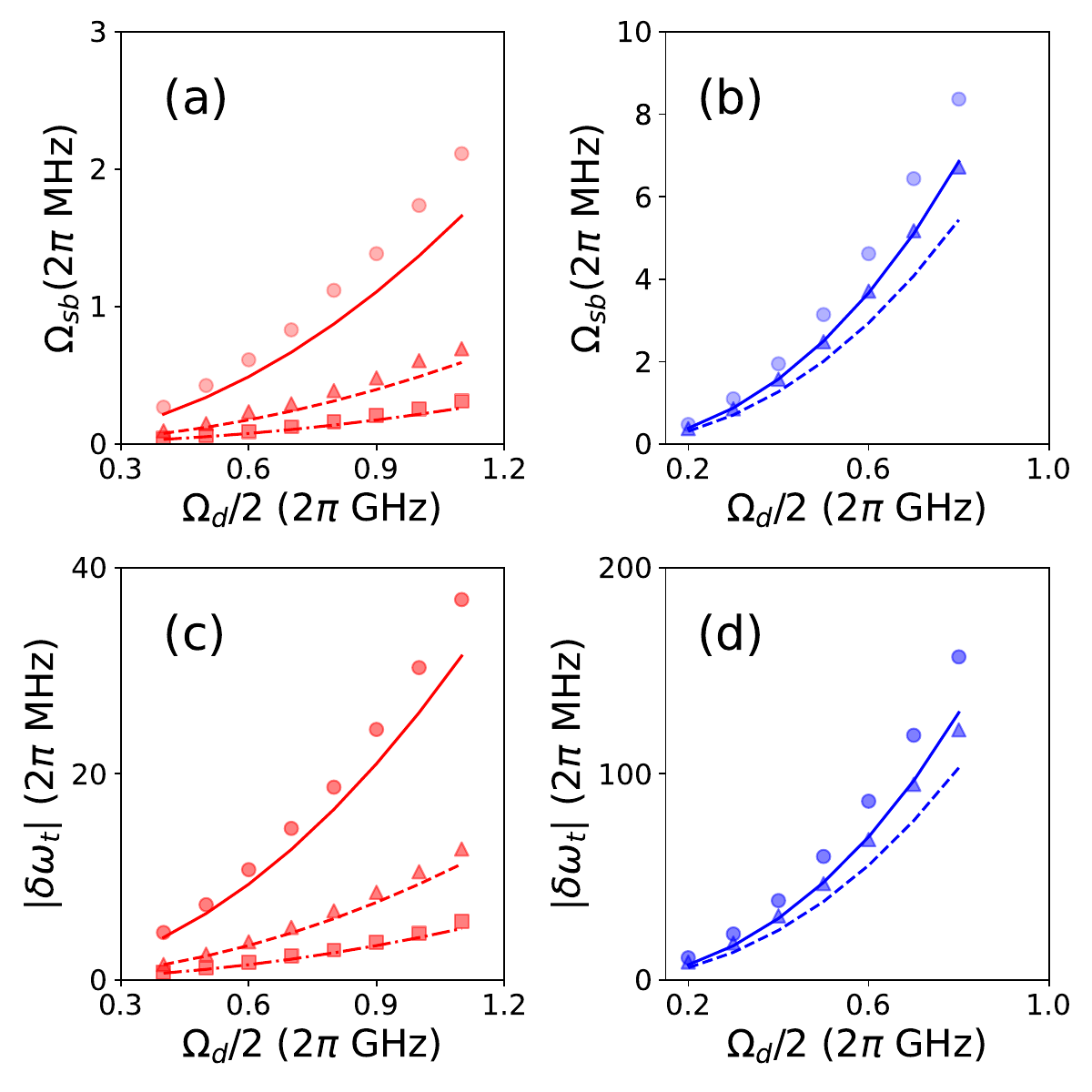}
    \caption{Comparison between analytical and numerical calculations. All of the contents in the figures are the same as Fig~\ref{fig:experiment} except that the lines are obtained based on Eq.~\ref{eq2}.}
    \label{fig:Sim_main_suppl}
\end{figure}

In this section, we confirm that Eq.~\ref{eq3-1} more accurately predicts the $\delta\omega_{t}$ and $\Omega_{sb}$ than Eq.~\ref{eq2}.
In Fig.~\ref{fig:Sim_main_suppl}, we compare the analytical calculation based on Eq.~\ref{eq2} and numerical simulation results in Fig.~\ref{fig:experiment}. We can clearly see the discrepancy between the analytical and numerical results becomes larger than that in Fig.~\ref{fig:experiment}.

\begin{figure}
    \centering
    \includegraphics[width=1\linewidth]{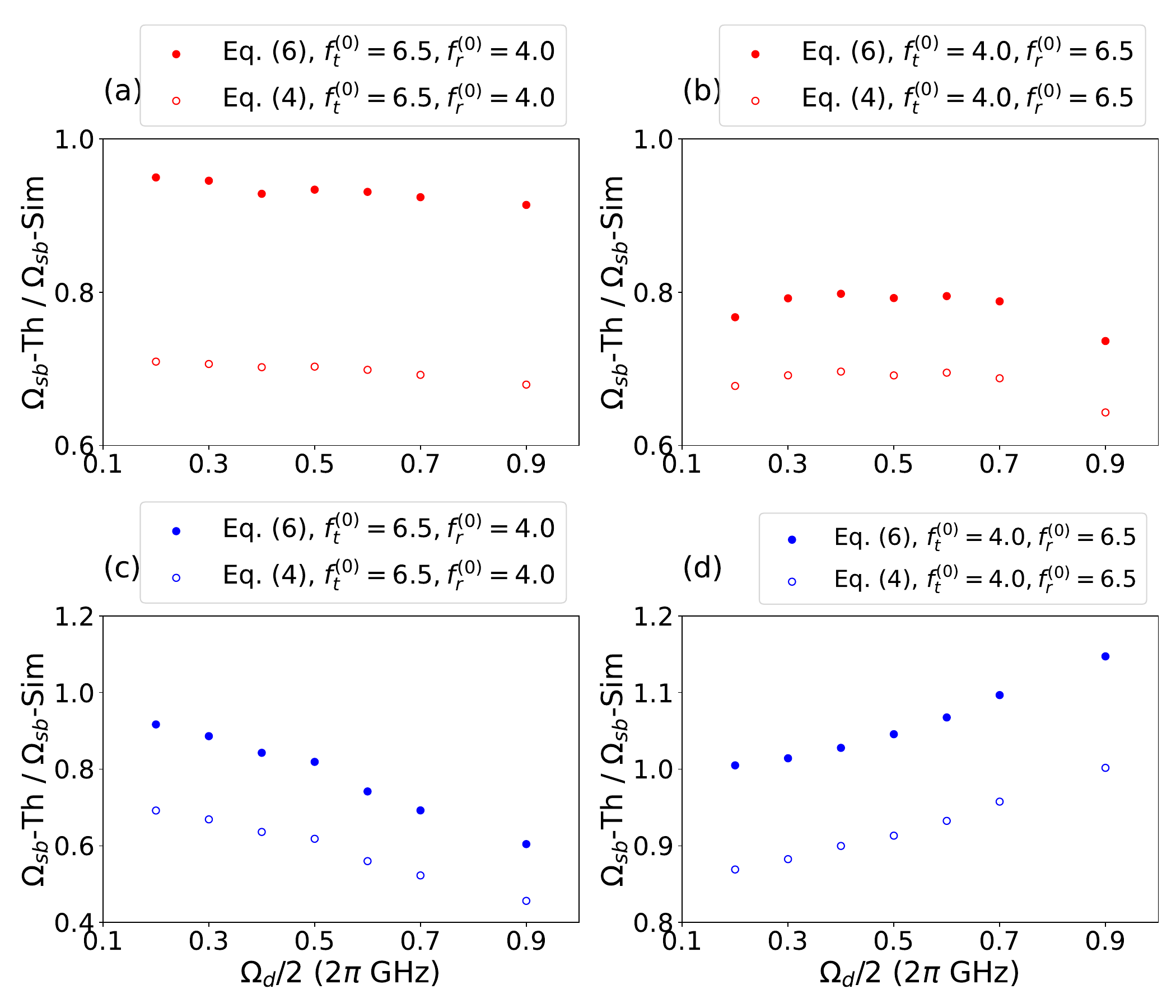}
    \caption{Additional simulation results ($\Omega_{sb}$-$\textup{Sim}$) and comparison to the analytical theory ($\Omega_{sb}$-$\textup{Th}$). (a-b) BS interaction. (c-d) TMS interaction.  Simulation is performed with four different system parameter combinations (see legend). $f_{t,r}^{(0)}$ are defined by $\omega_{t,r}^{(0)}/2\pi$. $g/2\pi$ in the simulation is 200 MHz. $f_{t,r}^{(0)}$ in (a,c) are 6.5 and 4.0 GHz, respectively. $f_{t,r}^{(0)}$ in (b,d) are 4.0 and 6.5 GHz, respectively.}
    \label{fig:Sim_summary}
\end{figure}

In Fig~\ref{fig:Sim_summary}, we perform the additional simulation with various system parameters and compare the numerically simulated sideband transition rates ($\Omega_{sb}$-$\textup{Sim}$) to the theoretical calculations ($\Omega_{sb}$-$\textup{Th}$). We compare two different theoretical approaches based on Eq.~\ref{eq2} and Eq.~\ref{eq3-1}, respectively. Aside from one case (Fig~\ref{fig:Sim_summary}-d), $\Omega_{sb}$-$\textup{Th}$ based on Eq.~\ref{eq3-1} are closer to $\Omega_{sb}$-$\textup{Sim}$. Even in Fig~\ref{fig:Sim_summary}-d, $\Omega_{sb}$-$\textup{Th}$ based on Eq.~\ref{eq3-1} is more accurate with low driving amplitudes.

\begin{thebibliography}{10}
\bibitem{note} 
Rotating wave approximation (RWA) is also available for time-independent Hamiltonian in the lab frame. In this case, dropping energy non-conservative terms constitutes the RWA. Nonetheless, we confine the scope of this work to when the breakdown of the RWA results from the time-periodical driving terms.
\end{thebibliography}
\end{document}